\documentclass[12pt,aps,prd,superscriptaddress,showpacs,longbibliography,floatfix,nofootinbib]{revtex4-1}

\usepackage[utf8]{inputenc}
\usepackage{slashed}
\pdfoutput=1

\usepackage{color}
\usepackage{graphicx}   
\usepackage{bm}
\usepackage{amsmath}
\usepackage{amsfonts}
\usepackage{eufrak}
\usepackage{hyperref}

\newcommand{\be}{\begin{equation}}
\newcommand{\ee}{\end{equation}}

\newcommand{\ba}{\begin{eqnarray}}
\newcommand{\ea}{\end{eqnarray}}

\begin{document}

\title{Out-Of-Time-Ordered-Correlators for the Pure Inverted Quartic Oscillator: Classical Chaos meets Quantum Stability}

\author{Paul Romatschke}
\affiliation{Department of Physics, University of Colorado, Boulder, Colorado 80309, USA}
\affiliation{Center for Theory of Quantum Matter, University of Colorado, Boulder, Colorado 80309, USA}

\begin{abstract}
Out-of-time-ordered-correlators (OTOCs) have been suggested as a means to diagnose chaotic behavior in quantum mechanical systems. Recently, it was found that OTOCs display exponential growth for the inverted quantum harmonic oscillator, mirroring the fact that this system is classically and quantum mechanically unstable. In this work, I study OTOCs for the inverted anharmonic (pure quartic) oscillator in quantum mechanics, finding only oscillatory behavior despite the classically unstable nature of the system. For higher temperature, OTOCs seem to exhibit saturation consistent with a value of $-2 \langle x^2 \rangle_T \langle p^2 \rangle_T$ at late times. I provide analytic evidence from the spectral zeta-function and the WKB method as well as direct numerical solutions of the Schr\"odinger equation that the inverted quartic oscillator possesses a real and positive energy eigenspectrum, and normalizable wave-functions.
\end{abstract}

\maketitle

\section{Introduction}

Out-of-time-correlators (OTOCs) in quantum mechanical systems have been of interest recently \cite{Hashimoto:2017oit,Hashimoto:2020xfr,Romatschke:2020qfr,Das:2020nay,Morita:2021syq,Sundar:2021lyv,Buividovich:2022jgv,Tucci:2022hmq,Li:2023ukv,Hashimoto:2023swv,Huang:2023qjo,Trunin:2023rwm,Huang:2023grt,Huang:2024tba}. For harmonic oscillator potentials, OTOCs are known to be purely oscillatory, while exponentially growing OTOCs have been found in classically chaotic systems such as the two-dimensional stadium billiard and non-linearly coupled oscillators \cite{Hashimoto:2017oit,Akutagawa:2020qbj}.

Ref.~\cite{Hashimoto:2020xfr} studied the system of an \textit{inverted} harmonic oscillator, finding that OTOCs display exponential growth with an exponent similar to  the classical Lyapunov exponent obtained for the classically unstable potential. 

In this work, I will study the \textit{inverted quartic} oscillator, which is also classically unstable. Naively, one could therefore expect to find exponential growth for the OTOCs as in the case of the inverted harmonic oscillator. On the other hand, owing to the fact that the inverted quartic oscillator is a special case of systems protected by symmetry \cite{Bender:1998ke}, one could also expect that OTOCs display oscillatory behavior and no exponential growth. The resolution of this tension provides the motivation for this work to study OTOCs for the inverted quartic oscillator.

\section{Out-of-time-ordered correlators in quantum mechanics}

This section largely follows the setup outlined in Ref.~\cite{Hashimoto:2017oit,Hashimoto:2020xfr}
which calculated microcanonical and thermal OTOCs in quantum mechanics. Denoting the quantum mechanical Hamiltonian as $H=H(\hat x,\hat p)$, the microcanonical OTOC for a fixed energy eigenstate $|n\rangle$ is given by \cite{Hashimoto:2017oit}
\begin{equation}
  c_n(t)=-\langle n| [x(t),p(0)]^2 |n\rangle\,.
\end{equation}
For a Hamiltonian of the form
\begin{equation}
  \label{hamiltonian}
  H=p^2+V(x)\,,
\end{equation}
the microcanonical OTOC may be expressed in terms of standard matrix elements as
\begin{equation}
  c_n(t)=\sum_{m=0}^\infty b_{nm}(t)b_{mn}(t)\,,\quad b_{nm}(t)\equiv -i \langle n| [x(t),p]|m\rangle\,,
\end{equation}
where
\cite{Hashimoto:2017oit} 
\begin{equation}
  b_{nm}(t)=\frac{1}{2}\sum_{k=0}^\infty x_{nk}x_{km}\left[\left(E_k-E_m\right) e^{-i t \left(E_k-E_n\right)}+\left(E_k-E_n\right)e^{i t \left(E_k-E_m\right)}\right]\,,\quad x_{nm}\equiv \langle n|x|m\rangle\,.
\end{equation}
The OTOC for system coupled to a heat-bath is then defined as
\begin{equation}
  O_T(t)\equiv-\langle[x(t),p(0)]^2\rangle_T\,,
\end{equation}
where $\langle \cdot \rangle_T$ denotes calculation of the expectation value at finite temperature $T$. The finite-temperature OTOC can therefore be related to the microcanonical OTOC as
\begin{equation}
  \label{otocdef}
  O_T(t)=Z^{-1}(\beta)\sum_{n=0}^\infty e^{-\beta E_n} c_n(t)\,,\quad Z(\beta)=\sum_{n=0}^\infty e^{-\beta E_n}\,,
\end{equation}
where $H|n\rangle=E_n|n\rangle$ and $Z(\beta)$ denotes the partition function of the system.

\section{Classical Instability of the Inverted Quartic Oscillator}

Let me now consider the Hamiltonian (\ref{hamiltonian}) with an inverted pure quartic potential $V(x)$ given by
\begin{equation}
  \label{eq:apot}
  V(x)=-x^{4}\,.
\end{equation}
The classical equations of motion for this potential then read
\begin{equation}
  \label{classicaleom}
\ddot{x}(t)=16 x^3(t)\,.
\end{equation}
It is straightforward to see that the classical system is unstable. For instance, an analytic solution to (\ref{classicaleom}) with $\dot{x}(0)=0$ and $x(0)=\frac{1}{4}$ is
\be
x(t)=\frac{1}{4{\rm cn}\left(t;\frac{1}{2}\right)}\,,
\ee
where ${\rm cn}$ denotes a Jacobi elliptic function, which can be thought of as a generalization of the cosine (oscillatory behavior). The Jacobi elliptic ${\rm cn}$ becomes zero at its quarter period given by the complete elliptic integral $t=K\left(\frac{1}{2}\right)\simeq 1.85$, so that $x(t=K\left(\frac{1}{2}\right))\rightarrow \infty$. As a consequence, a classical particle with initial position $x(0)=\frac{1}{4}$ reaches infinity in a finite time $t=K\left(\frac{1}{2}\right)$, indicating a super-exponential instability of the classical system.

Note that this instability is poorly captured by a Lyapunov exponent, which only captures exponential-in-time divergences.

\section{Quantum Stability of the inverted quartic oscillator}

Remarkably, despite this classical instability, the system described by the quantum mechanical equations of motion is stable. This is analogous to the hydrogen atom, where the electron solution is famously stable in quantum mechanics, despite being classically unconditionally unstable. A first indication of the stability of the inverted quartic oscillator was reported in Ref.~\cite{Bender:1998ke} by studying the energy-eigenspectrum $E_n$, $n=0,1,\ldots $of the Hamiltonian (\ref{hamiltonian}) with inverted potential (\ref{eq:apot}). Let me review what is known about $E_n$ analytically first by re-deriving results for sums of the form $\sum_n E_n^{-1}$, followed by the WKB analysis of the system, before presenting the numerical eigenvalues and eigenfunctions solving $H \psi_n(x) = E_n \psi_n(x)$.

\subsection{Spectral Zeta Functions}

For systems exhibiting a discrete and non-degenerate energy eigenspectrum $E_n$, the spectral zeta-function is defined as
\be
\zeta_S(s)=\sum_{n=0}^\infty \frac{1}{E_n^s}\,,
\ee
in analogy to the Riemann zeta-function. Remarkably, even if the energy eigenspectrum is unknown, closed-form analytic expressions for $\zeta_S$ can be derived \cite{Randall:1996}. For completeness, I will rederive the general formula for $\zeta_S$ in terms of the Green's function of an operator, before specializing to the case of (\ref{eq:apot}).

For a Hamiltonian of the form (\ref{hamiltonian}), the time-independent Schr\"odinger equation reads
\be
\label{schro}
-\psi^{\prime \prime}(x)+V(x) \psi(x)=E \psi(x)\,.
\ee
A systematic construction of the solution $\psi(x)$ may be constructed by expanding in powers of the energy $E$,
\be
\label{expansion}
\psi(x)=\sum_{m=0}^\infty \frac{E^m}{m!}\psi_{(m)}(x)\,,
\ee
and calculating $\psi_{(m)}(x)$ order-by-order in $E$ through solving (\ref{schro}). The corresponding recursion relations for $\psi_{(m)}(x)$ are given by
\be
L_x \psi_{(m)}(x)=m\psi_{(m-1)}(x)\,,\quad L_x\equiv -\partial_x^2+V(x)\,,
\ee
with the starting condition $L_x \psi_{(0)}(x)=0$ subject to the boundary conditions
\be
\label{bcs}
\psi_{(0)}(x\rightarrow -\infty)=C_-\,,\quad \psi_{(0)}^\prime(x\rightarrow -\infty)=D_-\,.
\ee
The recursion relations can be formally solved by introducing the Green's function for the differential operator $L$,
\be
L_x G(x,y)=\delta(x-y)\,,
\ee
so that
\be
\psi_{(m)}(x)=\psi^{\rm hom}_{(m)}(x)+m \int_{-\infty}^\infty dy G(x,y) \psi_{(m-1)}(y)\,,
\ee
with $L_x \psi^{\rm hom}_{(m)}(x)=0$. This homogeneous second-order differential equation possesses two linearly independent solutions $f_{\pm}(x)$, so that the Green's function can be written as
\be
G(x,y)=W^{-1}\left[f_+(x)f_-(y)\theta(x-y)+f_-(x)f_+(y) \theta(y-x)\right]\,,
\ee
with $W[f_+,f_-]=f_+ f_-^\prime-f_- f_+^\prime$ the Wronskian. Let me choose $f_+(x)$ to be well behaved near $x\rightarrow \infty$, in particular $f_+(x\rightarrow \infty)\rightarrow 0$. Using the form of the Green's function, one finds
\ba
\psi_{(m)}(x)=\psi^{\rm hom}_{(m)}(x)+ \frac{m f_+(x)}{W}\int_{-\infty}^x dy f_-(y) \psi_{(m-1)}(y)+\frac{m f_-(x)}{W}\int_x^\infty dy f_+(y) \psi_{(m-1)}(y)\,.
\ea
Since $\psi_{(0)}(x)$ fulfills the boundary condition near $x\rightarrow -\infty$, one must have  $\psi_{(m)}(x\rightarrow -\infty)=0$ for all $m\geq 1$, which fixes the homogeneous solution $\psi^{\rm hom}_{(m)}(x)$ above. Hence
\be
\psi_{(m)}(x)=\frac{m f_+(x)}{W}\int_{-\infty}^x dy f_-(y) \psi_{(m-1)}(y)-\frac{m f_-(x)}{W}\int_{-\infty}^x dy f_+(y) \psi_{(m-1)}(y)\,.
\ee
Plugging these results back into the expansion (\ref{expansion}), one finds that
\be
\psi(x\rightarrow \infty)=\psi_{(0)}(x)-f_-(x)\sum_{m=1}^\infty E^m \int_{-\infty}^\infty \frac{f_+(y)\psi_{(m-1)}(y)}{W (m-1)!}\,,
\ee
because $f_+(x\rightarrow \infty)\rightarrow 0$. Similarly, since $\psi_{(0)}(x)$ is a linear combination of $f_\pm(x)$, near $x\rightarrow \infty$ only the component proportional to $f_-(x)$ survives. One may thus choose $\psi_{(0)}(x)=f_-(x)$ and hence 
\be
\psi(x\rightarrow \infty)=f_-(x)\left[1-\sum_{m=1}^\infty E^m \int_{-\infty}^\infty \frac{f_+(y)\psi_{(m-1)}(y)}{W (m-1)!}\right]\,.
\ee
Typically, $f_-(x)$ is not well-behaved near $x\rightarrow \infty$, so in order to have a normalizable wave-function near $x\rightarrow \infty$ the coefficient multiplying $f_-(x)$ must vanish:
\be
F(E)\equiv 1+\sum_{m=1}^\infty E^m S_m=0\,, \quad S_m\equiv - \int_{-\infty}^\infty \frac{f_+(y)\psi_{(m-1)}(y)}{W (m-1)!}\,.
\ee
The function $F(E)$ is non-zero for arbitrary $E$, but the zeros of $F(E)$ correspond to the special values $E=E_n$ for which a normalizable solution $\psi(x)$ exists. Hence the zeros of $F(E)$ correspond to the eigenvalues of the Hamiltonian. Rewriting
\be
F(E)=\prod_{n=0}^\infty \left(1-\frac{E}{E_n}\right)\,,
\ee
one finds that the coefficients $S_m$ are related to the sums over eigenvalues $E_n$, specifically
\ba
\zeta_S(1)=\sum_{n=0}^\infty \frac{1}{E_n}&=&\frac{1}{W}\int_{-\infty}^\infty dy f_+(y) f_-(y)\nonumber\\
\zeta_S(2)=\sum_{n=0}^\infty \frac{1}{E_n^2}&=&\frac{2}{W^2}\int_{-\infty}^\infty dy f_+^2(y) \int_{-\infty}^y dz f_-^2(z)\nonumber\\
\zeta_S(3)=\sum_{n=0}^\infty \frac{1}{E_n^3}&=&\frac{3!}{W^3}\int_{-\infty}^\infty dy f_+^2(y) \int_{-\infty}^y dz f_-(z)f_+(z)\int_{-\infty}^z dw f_-^2(w)\,.
\ea

Therefore, knowledge of the homogeneous solutions $f_\pm(x)$ can be used to calculate the spectral zeta-function of the quantum mechanical problem. For the regular quartic oscillator with potential $V(x)=+x^4$, this has been studied in great detail in Ref.~\cite{voros1983return}, uncovering many unusual and unexpected properties. For instance, solving the homogeneous Schr\"odinger equation for $V(x)=+x^4$, one readily identifies
\be
V(x)=x^4:\ f_+(x>0)=\theta(x)\sqrt{x} K_{\frac{1}{6}}\left(\frac{x^3}{3}\right)+\theta(-x)\frac{\pi \sqrt{-x}}{2\sin\frac{\pi}{6}} \left[I_{-\frac{1}{6}}\left(\frac{-x^3}{3}\right)+I_{\frac{1}{6}}\left(\frac{-x^3}{3}\right)\right]\,,
\ee
where $K_\nu(x)$ denotes a modified Bessel function. Note that for $x<0$, $f_+(x)$ is given by the analytic continuation of $f_+(x)$ that is continuous and differentiable across $x=0$.

For large argument, this solution has the asymptotic expansion
\be
\label{fplusplus}
V(x)=x^4:\quad f_+(x\gg 1)\propto \frac{e^{-\frac{x^3}{3}}}{x}\,,
\ee
so it is well behaved for $x\rightarrow \infty$. Once $f_+(x)$ has been constructed, $f_-(x)$ is given by \cite{voros1983return}
\be
\label{zetamaster}
f_-(x)=f^*_+(-x)\,,
\ee
where it should be noted that for $V(x)=x^4$, $f^*_+(-x)=f_+(-x)$ because $f_+(x)$ is real.  The resulting functions $f_+,f_-$ can then be used to calculate the sum over eigenvalues of the quartic oscillator, e.g.
\be
\label{quartic}
V(x)=x^4:\quad \sum_{n=0}^\infty \frac{1}{E_n}=\frac{3\Gamma^5\left(\frac{1}{3}\right)}{12^{\frac{1}{3}}8\pi^2}\simeq 2.28991\,,
\ee
which matches the sum over the (only approximately known) eigenvalues $E_n$ \cite{Bender:1977dr,Hioe:1978jj,voros1983return}.

For the inverted quartic oscillator $V(x)=-x^4$, a solution bases for $f_\pm(x)$ is given in terms of $\sqrt{x}J_{\pm \frac{1}{6}}\left(\frac{x^3}{3}\right)$. To construct $f_+(x)$, I combine these solutions so that it matches the asymptotic form suggested by (\ref{fplusplus}), e.g.
\be
\label{fplusbc}
f_+(x\gg 1)\propto e^{-\frac{i x^3}{3}}\,.
\ee
Therefore,
\be
V(x)=-x^4:\ f_+(x>0)=\sqrt{x}\left[e^{\frac{i \pi}{12}}J_{-\frac{1}{6}}\left(\frac{x^3}{3}\right)-e^{-\frac{i \pi}{12}}J_{\frac{1}{6}}\left(\frac{x^3}{3}\right)\right]\,,
\ee
and using the constraints that $f_+(x)$ has to be continuous and differentiable at $x=0$ I get
\be
f_+(x<0)=\sqrt{-x}\left[e^{\frac{i \pi}{12}}J_{-\frac{1}{6}}\left(\frac{-x^3}{3}\right)+e^{-\frac{i \pi}{12}}J_{\frac{1}{6}}\left(\frac{-x^3}{3}\right)\right]\,.
\ee
The second solution $f_-(x)$ is once again fixed by (\ref{zetamaster}). Noting that $f_{\pm}(-x)=f_{\mp}^*(x)$ all integrals reduce to integrals over the positive real axis, and I find
\be
V(x)=-x^4:\quad \zeta_S(1)=\sum_{n=0}^\infty \frac{1}{E_n}=\frac{\Gamma^5\left(\frac{1}{3}\right)}{12^{\frac{1}{3}}4\pi^2}\simeq 1.52661\,,
\ee
matching the results reported in Refs.~\cite{mezincescu2000some,bender2001comment}. 
Curiously, this is equal to $\frac{2}{3}$ of the eigenvalues of the quartic oscillator given in (\ref{quartic}).
Furthermore, I find for $V(x)=-x^4$:
\ba
\zeta_S(2)=\sum_{n=0}^\infty \frac{1}{E_n^2}&=&2 \int_0^\infty dx dy \frac{f_+^2(x) f_+^{* 2}(y)}{W^2}+4 {\rm Re}\int_0^\infty dx \int_0^x dy \frac{f_+^2(x) f_-^2(y)}{W^2}\simeq 0.501049\nonumber\\
\zeta_S(3)=\sum_{n=0}^\infty \frac{1}{E_n^3}&\simeq & 0.31577\,,
\ea
where I remark that an analytic expression for $\zeta_S(2)$ in terms of generalized hypergeometric function is possible, but somewhat unenlightening. The results for $\zeta_S(1),\zeta_S(2),\zeta_S(3)$ can be compared to summing the numerically calculated eigenvalues given in Tab.~\ref{tab:one} and the WKB eigenvalues for $n\geq 5$ (\ref{enwkb}) given below, finding agreement up to numerical precision.

The above results for $\zeta_S(3)$ immediately leads to the following inequality
\be
E_0\geq \zeta^{-\frac{1}{3}}_S(3)\simeq 1.4685\,.
\ee
In addition, one may safely pose that eigenvalues scale as $\frac{E_n}{E_{n-1}}>\frac{n+\frac{1}{2}}{n-\frac{1}{2}}$ for $n\geq 1$. From this, the above inequality for $E_0$, and the spectral zeta functions, one may derive another inequality
\be
E_1\leq \left(\frac{\sum_{n=1}^\infty \frac{9}{(2n+1)^2}}{\zeta_S(2)-\zeta_S^{\frac{2}{3}}(3)}\right)^{\frac{1}{2}}\simeq 7.5\,,
\ee
so that together with $E_1\geq E_0$ one can bound the ground-state energy for the inverted quartic oscillator:
\be
1.4685 \leq E_0 \leq 7.5\,.
\ee
This result has the interpretation of a proof for the existence of a finite and positive ground state energy for the inverted quartic quantum oscillator.

\subsection{WKB analysis}

In quantum mechanics, the high-lying eigenstates may be captured by a semi-classical solution of the problem, also known as WKB analysis. This analysis has been performed in \cite{Bender:1998ke,Bender:2023cem}, and I repeat the main steps in my notation for completeness, emphasizing an unusual property of the WKB wave-functions for the inverted quartic oscillator.

For the WKB analysis of the potential $V(x)=-x^4$, one uses the ansatz $\psi(x)=e^{i S(x)}$ in (\ref{schro}), finding
\be
S^\prime(x)=\pm\sqrt{E+x^4+i S^{\prime\prime}(x)}\,.
\ee
Assuming $|S^{\prime\prime}|\ll E+x^4$ one can set up a systematic expansion for the solution. The lowest order leads to the solution
\be
S(x)=\pm \int_{x_-}^{x^+} dx \sqrt{E+x^4}\,,
\ee
where $x^{\pm}$ are the solutions of $E+x^4=0$. For classically stable potentials $V(|x|\rightarrow \infty)>0$, the points $x^\pm$ have the interpretation of real-valued 'turning-points' of a classical particle inside $V(x)$. However, as pointed out above, there are no classically stable solutions for the inverted quartic oscillator. Nevertheless, it has been shown that one can \textit{formally} use the WKB method by allowing $x^\pm$ to be complex, specifically $x^+=e^{-\frac{i \pi}{4}}E^{\frac{1}{4}}$, $x^-=e^{-\frac{3 i \pi}{4}}E^{\frac{1}{4}}$ \cite{Bender:2023cem}. The Sommerfeld quantization condition
\be
\int_{x_-}^{x^+} dx \sqrt{E+x^4}=\left(n+\frac{1}{2}\right)\pi\,,
\ee
then leads to the WKB eigenvalues
\be
\label{enwkb}
E_n^{\rm WKB}=\left(\frac{3\pi \left(n+\frac{1}{2}\right)}{4 \cos\frac{\pi}{4} K(-1) }\right)^{\frac{4}{3}}\,,
\ee
where $K(-1)$ again denotes the complete elliptic integral of the first kind. In addition, one finds the WKB wave-functions
\be
\label{psiwkb}
\psi_n^{\rm WKB}(x)=k_n \frac{e^{-\frac{i}{3}\left(x \sqrt{E_n+x^4}-2 e^{\frac{3 i \pi}{4}}E_n^{\frac{3}{4}}F\left({\rm atan}\left(e^{\frac{i\pi}{4}}E_n^{-\frac{1}{4}}x\right);2\right)\right)}}{(E_n+x^4)^{\frac{1}{4}}}\,,
\ee
where $k_n$ are real-valued normalization constants, $F(\phi,m)$ denotes the incomplete elliptic integral of the first kind of modulus $m$, and I have used the boundary condition (\ref{fplusbc}) to fix one of the integration constants. In stark contradistinction to the WKB wave-functions encountered for classical stable potentials, note that $\psi_n^{\rm WKB}(x)$ are well behaved for all $x\in \mathbb{R}$, and obey $\psi_n^{\rm WKB}(|x|\rightarrow \infty)\rightarrow 0$, cf. Fig.~\ref{fig:fig2}.  I will show below that the WKB values $E_n$ and wave-functions $\psi_n^{\rm WKB}$ are good approximations for the 'exact' eigenvalues $E_n$ and eigenfunction solutions of the Schr\"odinger equation for $n\gg 1$.

The WKB wave-functions for the inverted quartic oscillator fulfill the property
\be
\psi_n^{\rm WKB}(x)=\left(\psi_n^{WKB}(-x)\right)^*={\cal PT}\psi_n^{\rm WKB}(x)\,,
\ee
e.g. they are symmetric under both parity ${\cal P}$ and complex conjugation ${\cal T}$. By contrast, the wave-functions for the usual (positive-sign potential) quartic oscillator fulfill $\psi_n^\dagger(x)=\psi_n(x)$. This suggests that the inner product for the Hilbert space for the inverted quartic oscillator differs from the usual definition, because the symmetry is different \cite{Bender:2023cem}. Numerical integration gives
\be
\int_{-\infty}^\infty dx  \left[{\cal PT}\psi_n^{\rm WKB}(x)\right] \psi_n^{\rm WKB}(x)=\int_{-\infty}^\infty dx \left[\psi_n^{\rm WKB }(x)\right]^2\propto (-1)^n |k_n|^2\,,
\ee
so that a positive-norm inner product requires
\be
\langle n^{\rm WKB}|n^{\rm WKB}\rangle\equiv (-1)^n \int_{-\infty}^\infty dx \left[{\cal PT}\psi_n^{\rm WKB}(x)\right] \psi_n^{\rm WKB}(x)\,.
\ee
I will employ a similar inner product to normalize the numerically calculated wave-functions for the inverted quartic oscillator below.

It should be pointed out that the rapidly oscillating nature of $\psi_n^{\rm WKB}(x)$ makes precision numerical integration on the real axis challenging, particularly for higher values of $n$. An easy remedy of the situation is provided by contour-deforming the integration path into the complex plane, which does not change the value of the integral. In particular, deforming the integration contour into a 'cone'
\be
\label{cone}
x=x(s)=s \left(e^{i\phi}\theta(-s)+e^{-i\phi}\theta(s)\right)\,,
\ee
with $s\in \mathbb{R}$ parametrizing the integration path and $\phi\in(0,\frac{\pi}{4})$ a fixed angle, and closing back to the real axis at $|x|\rightarrow \infty$ gives high-precision numerical results for any $n$. In this manner, I report the first few normalization constants $k_n$ obeying $\langle n^{\rm WKB}|n^{\rm WKB}\rangle=1$  in Tab.~\ref{tab:one}

\begin{table}
\begin{tabular}{|c|ccccccc|}
  \hline
  n & 0 & 1 & 2 & 3 & 4 & 5 & 6\\
  \hline
  $E_n$ & 1.4771498 & 6.003386 & 11.80243 & 18.4588 & 25.7918 & 33.6943 & 42.094\\
  $E_n^{\rm WKB}$ & 1.37651 & 5.9558 & 11.769 & 18.4321 & 25.7693 & 33.6747 & 42.0762\\ 
  $k_n$ & 2.22255 & 15.5198 &  88.7167 & 477.92 & 2501.34 & 12869.7 & 65019\\
  \hline
\end{tabular}
\caption{\label{tab:one} Numerical results for the WKB energies $E_n^{\rm WKB}$ and wave-function normalization constants $k_n$, cf. Eqns.~(\ref{enwkb}), (\ref{psiwkb}). For comparison, the numerically calculated ``exact'' eigenvalues $E_n$ are also reported.}
\end{table}

\subsection{Numerical calculation of Eigenvalues and Eigenfunctions}

\begin{figure*}[t]
  \includegraphics[width=\linewidth]{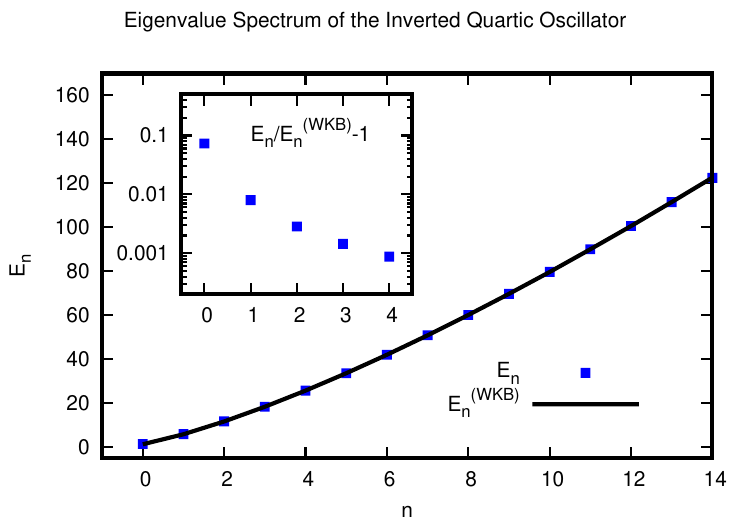}
  \caption{\label{fig:fig1} Eigenvalue spectrum $E_n$ of the inverted quartic oscillator, compared to the WKB approximation as a function of $n$. Inset shows $E_n/E_n^{\rm (WKB)}-1$ to demonstrate that eigenvalues rapidly converge to the WKB values (note the logarithmic scale).}
\end{figure*}

\begin{figure*}[t]
  \includegraphics[width=\linewidth]{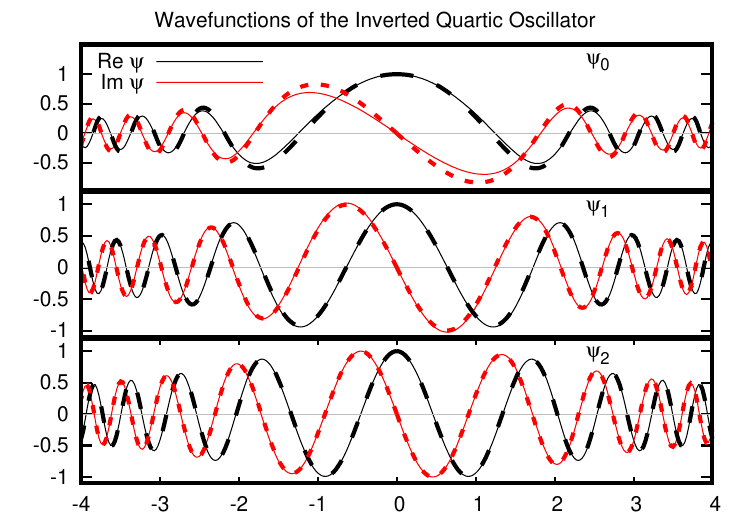}
  \caption{\label{fig:fig2} Eigenfunctions $\psi_0,\psi_1,\psi_2$ of the inverted quartic oscillator (full lines),  compared to the WKB wave-functions (dashed lines) as a function of $x$. Wave-functions have both real and imaginary parts, and are node-less.}
\end{figure*}

The WKB analysis above only provides approximate results for the eigenvalues and eigenfunctions that solve (\ref{schro}) for the inverted quartic oscillator with $V(x)=-x^4$. Numerous numerical methods exist to solve the Schr\"odinger equation adapted to different boundary conditions for $\psi(x)$. In the following, the boundary condition for the wave-functions is the same as for the WKB solutions above, namely I impose
\be
\psi_n(-x)=\psi_n^*(x)\,,
\ee
e.g. I demand that the wave-functions are also eigenfunctions of ${\cal PT}$. A straightforward way to obtain solutions to the Schr\"odinger equation is then as follows: first note that for numerical stability, it is convenient to solve the Schr\"odinger equation on the ``cone'' (\ref{cone}). Choosing in particular $\phi=\frac{\pi}{6}$, the Schr\"odinger equation turns into two copies (``+'' and ``-'' in the following):
\be
-\psi_{\pm}^{\prime\prime}(s)+s^4 \psi_{\pm}(s)=E_{\pm}\psi_{\pm}(s)\,,
\ee
with $E_{\pm}=e^{\mp \frac{i\pi}{3}}E$ the (complex-valued) energies for the Schr\"odinger equation originating from the positive/negative part of the real-axis. Note that for this choice of contour-angle, each copy of the Schr\"odinger equation formally corresponds to the regular quartic oscillator, with positive-definite potential. The two copies of the Schr\"odinger equation are subject to the condition that the wave-functions $\psi_n(x)$ on the real axis must be continuous and differentiable, e.g. $\psi_n(0^-)=\psi_n(0^+)$, $\psi_n^\prime(0^-)=\psi_n^\prime(0^+)$. Using contour-deformation (\ref{cone}), these conditions turn into conditions for $\psi_\pm$ at the origin. Specifically, changing sign of the real-valued contour parameter $s$ for $\psi_-$ such that $s>0$ for both $\psi_{\pm}$, one has
\be
\label{psibc1}
\psi_{-}(0)=\psi_+(0)\,,\quad \psi_-^\prime(0)=e^{-\frac{2 i \pi}{3}}\psi_+^\prime(0)\,.
\ee
as well as $\psi_\pm(s\rightarrow \infty)\rightarrow 0$. To solve this system, I write $\psi_\pm(s)=e^{-\frac{s^3}{3}}u_\pm(s)$, and then use the power-series ansatz for $u_\pm(s)$ from Ref.~\cite{bay1997spectrum} (see also appendix A in \cite{Romatschke:2023ztk}) to generate the wave-functions.
The boundary conditions (\ref{psibc1}) then lead to the numerical values $E_n$ and wave-functions $\psi_n(x)$. The comparison to the results from the WKB approximation for the eigenvalues can be found in Tab.~\ref{tab:one} and Fig.~\ref{fig:fig1}, and for the wave-functions in Fig.~\ref{fig:fig2}.

This comparison shows that both the energy eigenvalues and wave-functions become very well approximated by the WKB approximation for $n\gtrsim 3$. Nevertheless, for precise numerical calculations of expectation values the small difference between WKB and exact wave-functions becomes important because of numerical cancellations, so that in practice the numerically calculated wave-functions must be used. To give an example, recall that for the inverted quartic oscillator, the inner product is different from the usual Hermitian inner product. Specifically, one defines the inner product between energy eigenstates as
\be
\label{innerfull}
\langle n | m \rangle = \int dx (-1)^n {\cal P T}\psi_n(x) \psi_m(x)=\int dx (-1)^n \psi^*_n(-x) \psi_m(x)=\delta_{nm}\,,
\ee
which is easily verified using the numerically calculated wave-functions $\psi_n$ (but violated by the WKB wave-functions, which are not orthogonal). Since the wave-functions $\psi_n$ are orthogonal and complete \cite{Bender:2023cem}, the inner product generalizes to any superposition of wave-functions  $f(x)=\sum_n f_n \psi_n(x)$, $g(x)=\sum_n g_n \psi_n(x)$ with complex coefficients $f_n,g_n$ so that
\be
\langle f|g \rangle = \sum_{n,m}f^*_n g_m \int dx (-1)^n \psi^*_n(x) \psi_m(x)=
\sum_{n} f_n^* g_n =\left(\langle g|f \rangle \right)^*\,.
\ee

As an application, and to highlight the difference with respect to naive expectations, one can numerically calculate the expectation value of the kinetic energy, finding
\be
\label{v1}
\langle n  | p^2 | n \rangle = \frac{2}{3}E_n\,,
\ee
in accordance with the quantum virial theorem for a quartic potential \cite{Romatschke:2020qfr}. However, this implies -- and is confirmed by the direct numerical calculation -- for the expectation value of the potential
\be
\label{virial}
\langle n | x^4 | n \rangle = -\frac{1}{3}E_n\,,
\ee
e.g. a \textit{negative} expectation value for the operator $x^4$, see Ref.~\cite{Bender:2023cem} for a discussion on the physical interpretation of this finding. I close this paragraph by noting that despite the similarity between the WKB wave-functions and $\psi_n$, the corresponding WKB expectation value is given by
\be
\langle n^{\rm WKB} | x^4 | n^{\rm WKB} \rangle \simeq -E_n^{\rm WKB}\,,
\ee
meaning that it differs considerably from the result obtained from the virial theorem (\ref{virial}). Because of this sizable difference, the WKB wave-functions cannot be used to accurately calculate the OTOCs.

\section{Numerical Results for the OTOCs}

Having calculated the eigenvalues and eigenfunctions of the inverted quartic oscillator in the previous section, the OTOCs discussed in section II may be calculated numerically. The only difference with respect to the calculation for the usual (non-inverted) quartic oscillator \cite{Romatschke:2020qfr} is that the matrix elements $x_{nm}=\langle n | x |m\rangle$ do not fulfill $x_{mn}=x_{nm}^*$ because of the nature of the inner product (\ref{innerfull}). However, the matrix elements fulfill $x_{mn}=\pm x_{nm}^*$, so that the microcanonical OTOCs $c_n(t)$ are guaranteed to be real (though no longer positive-definite).

For the numerical evaluation of the $c_n(t)$, I have first checked that my numerical results for the wave-functions $\psi_n(x)$ are stable up to a given $n=K$. Numerical stability for the wave-functions is harder to achieve than for the regular quartic oscillator \cite{Romatschke:2020qfr}, so that in practice I achieve stability for wave-functions up to (including) $K=8$. I then calculate the matrix elements and OTOCs using energy levels up to $n<K$, and check for numerical stability of results as $n\rightarrow K$. Only results for quantities where numerical stability is found are reported in the following.

\begin{figure*}[t]
  \includegraphics[width=\linewidth]{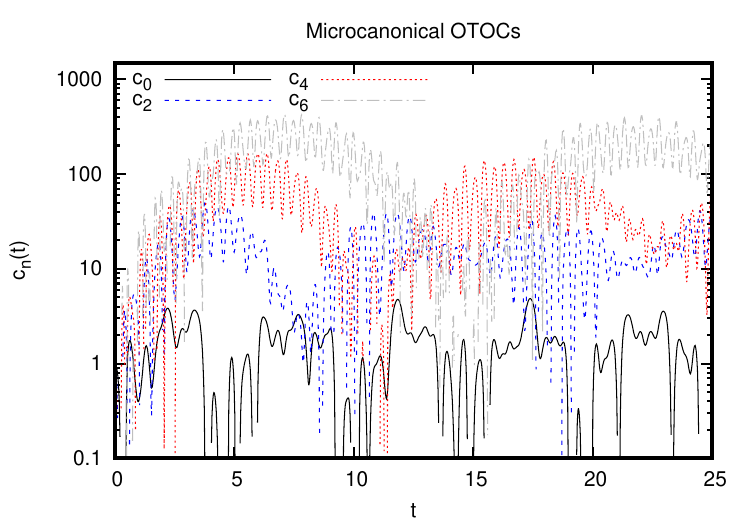}
  \caption{\label{fig3} Microcanonical OTOCs $c_n(t)$ as a function of time for $n=0,2,4,6$ (OTOCs for odd $n$ are qualitatively similar). }
\end{figure*}

\begin{figure*}[t]
  \includegraphics[width=\linewidth]{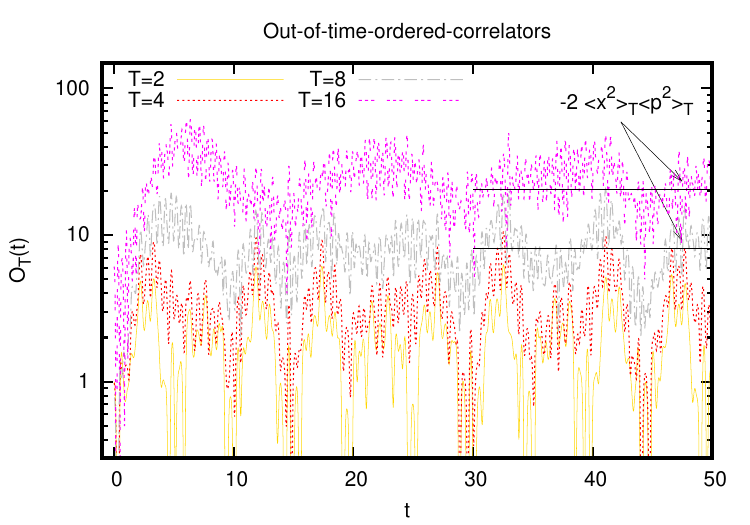}
  \caption{\label{fig:fig4} OTOCs $O_T(t)$ as a function of time for various temperatures $T$. For higher temperatures, there is a rapid rise at early times followed by a plateau-like behavior at late times. Note that this rapid rise is not exponential, but rather power-law like. For comparison, (\ref{qcs}) with negative sign is shown for $T=8,16$.}
\end{figure*}

Examples for low-lying microcanonical OTOCs $c_n(t)$ are shown in Fig.~\ref{fig3}, and (apart from the fact that they are not positive-definite) are qualitatively similar to the microcanonical OTOCs for the regular quartic oscillator \cite{Romatschke:2020qfr}. For higher $n$, $c_n(t)$ exhibit a rapid rise at early times, followed by a qualitatively similar behavior: a constant plus two harmonics,
again qualitatively similar to the regular quartic oscillator.

The resulting OTOC (\ref{otocdef}) for various temperatures is shown in Fig.~\ref{fig:fig4}. Thermal averaging of the microcanonical $c_n(t)$ seems to have the effect of further reducing the amplitude of the harmonic contributions such that at temperatures $T\geq 8$, Fig.~\ref{fig:fig4} suggests a rapid early-time rise followed by a plateau at late times for $O_T(t)$. While the early-time rise is clearly not a simple exponential, this qualitative behavior of $O_T(t)$ (rapid rise, saturation) has been associated with quantum chaotic behavior in systems that exhibit chaos, cf. Ref.~\cite{Akutagawa:2020qbj}. However, the same behavior has been found for the regular quartic oscillator, which is not expected to display quantum chaos \cite{Romatschke:2020qfr}.

For quantum chaotic system, it is expected that at late times \cite{Akutagawa:2020qbj}
\begin{equation}
  \label{qcs}
\lim_{t\rightarrow \infty}  O_T(t)=2\langle x^2\rangle_T \langle p^2\rangle_T\,,
\end{equation}
where $\langle {\cal O}\rangle_T$ again denotes the thermal expectation value of the operator ${\cal O}$. Using the quantum virial theorem, it is straightforward to calculate the expectation value of the momentum-operator squared, see Eq.~(\ref{v1}), whereas the expectation value of the position operator is found to be
\be
\langle n| x^2 |n \rangle<0\,,
\ee
which in light of Eq. (\ref{virial}) is not unexpected. The results for the $O_T(t)$ shown in Fig.~\ref{fig:fig4} indicate a plateau at late times with a positive value, and hence OTOCs are not described by (\ref{qcs}) at late times. However, it is curious to note that evaluating (\ref{qcs}) with \textit{negative sign} apparently describes the plateau seen in $O_T(t)$ for higher temperatures and late times, cf. Fig.~\ref{fig:fig4}, though more numerics for higher temperatures would be needed to corroborate this finding.

\section{Summary and Conclusions}

In the present work, I have studied OTOCs for the pure inverted quartic oscillator in quantum mechanics. I provided evidence based on the spectral zeta-function, the WKB method, as well as numerical solutions of the Schr\"odinger equation that the inverted quartic oscillator possesses a real and bounded energy eigenspectrum, and normalizable wave-functions, in agreement with earlier results \cite{Bender:1998ke}. This is in strong contrast to the inverted harmonic oscillator, which was studied in \cite{Hashimoto:2020xfr}.

The bounded energy spectrum and wave-functions for the inverted quartic oscillator then were used to calculate the microcanonical and thermal OTOCs for this quantum system, finding results that were in qualitative agreement with the regular quartic oscillator studied in Ref.~\cite{Romatschke:2023ztk}. In particular, there was no indication of exponential growth in the OTOCs for the inverted quartic oscillator for the temperatures studied. Again, this is in strong contrast to the inverted harmonic oscillator which displayed exponential growth \cite{Hashimoto:2020xfr}.

The absence of exponential growth in the OTOCs found in this work indicate that the inverted quartic oscillator is a system that -- while classically unstable -- is quantum mechanically stable. This finding may have wide-reaching consequences, which should be studied in follow-up work in the future.

For instance, the inverted quartic oscillator is part of a broader class of quantum systems possessing so-called ${\cal PT}$-symmetry, with potentials of the form $V(x)=-(i x)^{2+\alpha}$, with real $\alpha \geq 0$ \cite{Bender:1998ke}. Do all of these potentials lead to OTOCs that oscillate rather than show exponential growth?

Another direction to explore would be the behavior of OTOCs in quantum field theories with inverted quartic potentials, e.g. $V(\phi)=-\phi^4$ \cite{Bender:2015uxa}. Recognized as asymptotically free in 3+1 dimensions \cite{Symanzik:1973hx,Kleefeld:2005hf,Romatschke:2022jqg}, it has been pointed out recently that quantum triviality proofs \cite{Aizenman:2019yuo} do not apply to to inverted potentials \cite{Romatschke:2023ogd}. Inverted potentials of this form naturally occur in the continuum limit of renormalized scalar field theory in 3+1d in the large N limit\cite{Parisi:1975im,Romatschke:2023sce,Weller:2023jhc,Romatschke:2024yhx}, which is relevant for the Electroweak Sector of the Standard Model \cite{Romatschke:2024hpb}. Inverted potentials may also be relevant for the understanding of gauge theories \cite{Bender:2005zz}, notably QCD \cite{Romatschke:2022llf}.

An intermediate direction may be the study of inverted quartic potential field theory in lower dimensions \cite{Bender:2018pbv,Ai:2022csx,Lawrence:2023woz}, where direct numerical techniques \cite{Lawrence:2022afv,Romatschke:2023fax} could be used to study OTOCs.

  \section*{Acknowledgments}

  This work was partially supported by the Department of Energy, DOE award No DE-SC0017905.

\bibliography{otoc}
\end{document}